# Present and future constraints on top EW couplings

François Richard

Laboratoire de l'Accélérateur Linéaire, IN2P2–CNRS et Université de Paris–Sud, Bât. 200, BP 34, F–91898 Orsay Cedex, France.

**Abstract:** Various BSM models predict large deviations of the top electroweak couplings. Many examples will be given emphasizing the need for precise and model independent measurements of these couplings. Here I address the question of compatibility between theoretical predictions and the LEP/SLC/Tevatron constraints which, through loop effects, allow setting very significant constraints on these predictions. A Randall Sundrum inspired prediction will be discussed in more details. Finally the prospects offered by LHC and ILC measurements for top EW couplings are briefly reviewed.

*Talk given at Workshop on Top physics at the LC 5-6 March 2014 LPNHE Paris*

## I  Introduction

The SM model provides a consistent theory without UV divergences in higher order corrections. BSM models can modify the couplings of the Higgs boson to vector bosons and also the electroweak couplings of fermions to bosons. These modifications will induce UV divergences meaning that one needs to introduce an UV cut-off $\Lambda$ in the theory which can be interpreted as a mass scale at which the new physics sets in. As will be shown in the following, the appearance BSM deviations on couplings has some observable consequences on the $\epsilon_1$/T parameter measured at LEP/SLC and therefore allows to constrain these deviations once $\Lambda$ is fixed. After discussing an explicit example related to Higgs coupling to ZZ/WW, I will examine in more detail the occurrence of such deviations for top couplings to Z and W. This emphasis on the top and Higgs couplings relies on the generic idea that they are intimately related to the EWSB mechanism and therefore likely to be connected to similar BSM models. Well known such models are the Randall Sundrum[1], Little Higgs[2] and related composite theories. These models either assume mixing between the top quark and new heavy fermions or mixing between the Z/W bosons with some new heavy vector state which preferentially couple to top quarks and can induce significant variations of EW couplings. I will recall how the variables $S$/T and $\delta_b$ can be related to these variations and review the consequences for the models proposed on the market which assume that the top carries a great deal of compositeness.

ILC provides an ideal set up for this type of scenario, with a large set of observables allowed by beam polarisation, energy scans, cleanness of reconstruction and very high accuracy. LHC and Tevatron can also observe this type of effect if the occurrence of new vector states has a colored counter-part in the form of new heavy gluons predominantly coupled to top quarks. Here I will not address this possibility but only discuss the prospects to observe EW couplings at LHC by measuring the ttZ and the single top cross sections.

## II The Higgs case

To illustrate the method described in the introduction, let us take the example of Higgs coupling to bosonic pairs WW or ZZ. If this coupling deviates from the SM by a factor $\kappa_V$ different from 1, a logarithmic divergent term ~( $\kappa^2_V$ -1)log$\Lambda$ appears in the expression of T (and S) which allows to set a bound | $\kappa^2_V$ -1|<0.1 assuming that $\Lambda$ ~3 TeV. This bound is already very tight and shows that we need a very precise measurement of $\kappa_V$, at the % level, to observe a significant deviation. One should however presumably relax this bound to take into account the extra possible BSM contributions to S,T. Explicitly one has:

$$T = -\frac{3}{16\pi c_W^2}(1-\kappa_V^2)\log\frac{\Lambda^2}{M_H^2} \quad S = \frac{1}{12}(1-\kappa_V^2)\log\frac{\Lambda^2}{M_H^2}$$

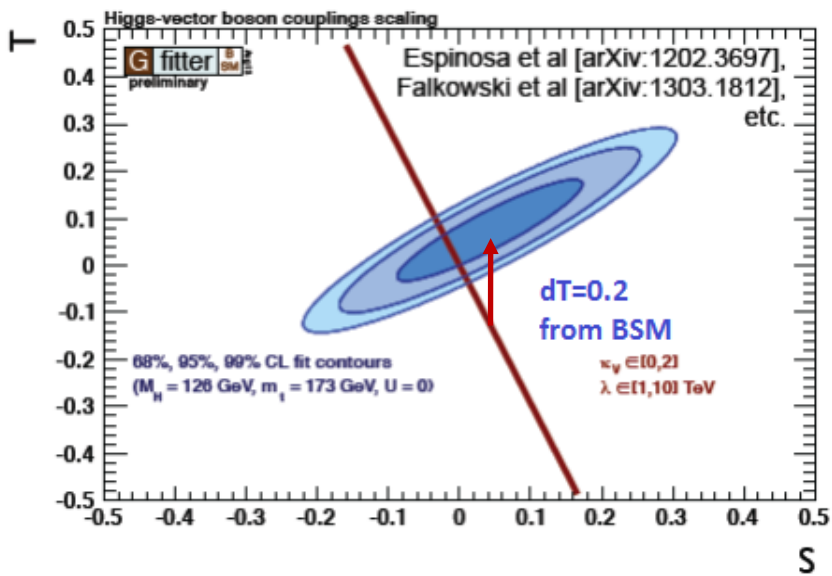

**Figure 1 Plot showing how the effect of Kv deviation (red diagonal line) can be compensated by BSM contribution to T,S.**

Figure 1 shows, for a fixed value of $\Lambda$, the correlated variations of S and T when $\kappa_V$ varies. Taking $\kappa^2_V$ =0.7 one sees that the corresponding solution falls outside the ellipses deduced with Gfitter[3] but one also sees that, assuming a positive contribution to T from the BSM model $T_{BSM}$=0.2, one can perfectly restore the agreement. This type of situation is likely to occur in composite models[4] and illustrates the methodology that could be followed when a more precise value of $\kappa_V$ will become available.

Why should $\kappa_V$ differ from 1? As we said in the introduction, this could occur within composite scenarios but not only. As an example, one of the many motivations to modify the Higgs potential is the possibility to generate an EW 1st order phase transition in the early universe and this job is not achieved by the SM alone. Reference [5] points out that, generically, if one adds some necessary ingredients to the SM, they should induce deviations in Higgs couplings. This is just one example among many which advocate for BSM extensions to explain our observed universe with DM and large matter anti-matter asymmetry.

## III The top quark case

### 3.1 Predictions

Top quarks being the heaviest fermions are the usual suspects for composite theories (or almost equivalently Randall Sundrum and Little Higgs) which predict deviations of the top couplings. Figure 2 shows the wide spectrum of predictions for these deviations in various models. References [6] to [13] provide the details on these predictions. This picture clearly demonstrates the motivation to separate the Ztt couplings into their left and right components which can be well achieved using the ILC set up with polarized beams (see section 5.2 on ILC).

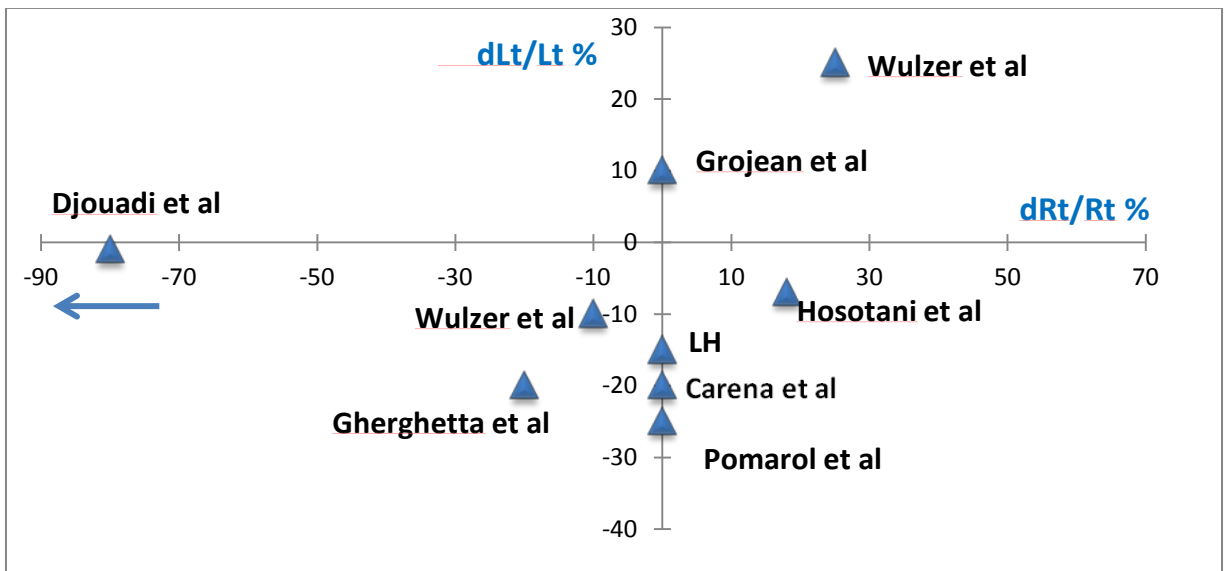

**Figure 2: Plot showing the predicted deviations of Z couplings to tL and tR in %. The Djouadi et al prediction falls outside of the scale.**

Note that LEP1 provides an indication for deviations on Rb and AFBb interpretable [7] as mainly due to ZbRbR with dbR/bR~0.2. This suggests the possibility of larger deviations on ZtRtR but obviously this varies with the models as discussed in 4.1. For the coupling ZbLbL one observes an almost negligible effect but this does not mean an absence of effect for ZtLtL as one could naively infer from the fact that bL and tL belong to the same EW doublet. Actually [14] have shown that with a custodial symmetry one can have full cancellation for ZbLbL while having enhanced effect for ZtLtL. Therefore fully deciphering the behavior of tL and tR is of outmost importance to understand the underlying structure and this can only be done at ILC with polarized beams. From the ttZ cross section measurable at LHC one cannot separate V and A (or equivalently L and R) components and the accuracy is quite limited (10% level).

## 3.2 Constraints

### 3.2.1 Loop predictions

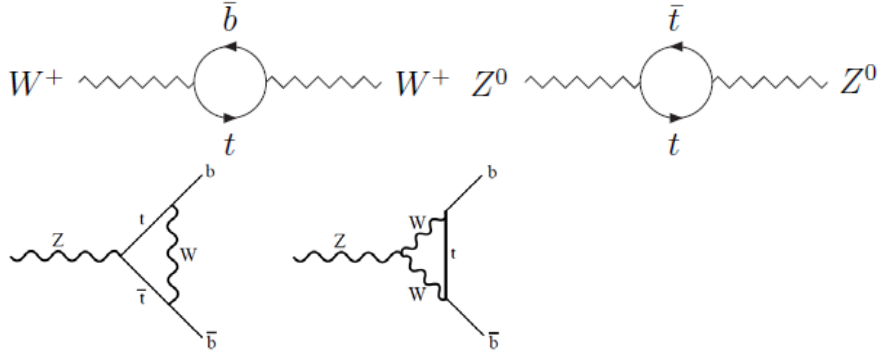

Above diagrams illustrate the connection of LEP/SLC measurements to top EW couplings.
Figure 3 from Snowmass2005 [15] illustrates this feature and I will explain shortly the origin of the two narrow bands which already provide constraints far better than the ones expected from LHC even at a luminosity of 3000 fb-1. The tight domain allowed for possible deviations on axial and vector couplings of Ztt obviously sounds discouraging but, for reasons already stated, need to be considerably relaxed considering the possible BSM contributions to $\epsilon_1$/T and $\epsilon_b$.

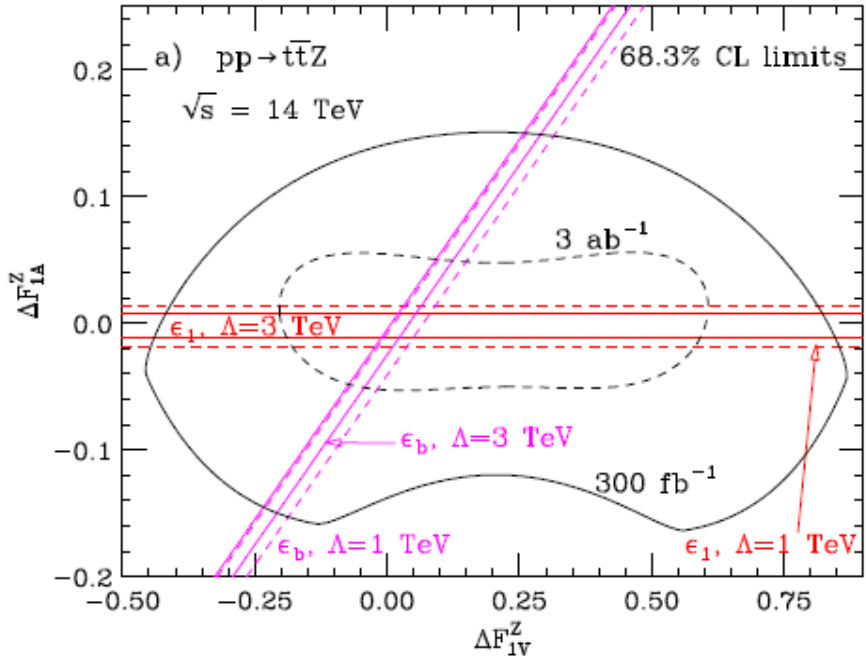

**Figure 3**: Plot extracted from Snowmass2005 which shows the measurement accuracies expected from LHC for 300 fb-1 (full contour) and from HL-LHC (dotted contour). The horizontal red lines correspond to the constraints from $\epsilon_1$, full line for $\Lambda$=3 TeV and dotted line for $\Lambda$=1 TeV and the diagonal lines are deduced from $\epsilon_b$.

From [16], one has:

$$\delta\epsilon_1 = \frac{3m_t^2 G_F}{2\sqrt{2}\pi^2}\left[\kappa_R^{NC} - \kappa_L^{NC} + \kappa_L^{CC} - \left(\kappa_R^{NC}\right)^2 - \left(\kappa_L^{NC}\right)^2 + \left(\kappa_L^{CC}\right)^2 + 2\kappa_R^{NC}\kappa_L^{NC}\right]\ln\frac{\Lambda^2}{m_t^2}$$

$$\delta\epsilon_b = \frac{m_t^2 G_F}{2\sqrt{2}\pi^2}\left(\kappa_L^{NC} - \frac{1}{4}\kappa_L^{NC}\right)\left(1 + 2\kappa_L^{CC}\right)\ln\frac{\Lambda^2}{m_t^2}$$

where $\delta\epsilon_1$ and $\delta\epsilon_b$ are the deviations due to the loop contributions from top quarks measured at

LEP1/SLC and where the various deviations on top couplings as defined in the following Lagrangian:

$$\mathcal{L} = \frac{g}{2c_W}\left(1 - \frac{4s_W^2}{3} + \kappa_L^{NC}\right)\overline{t_L}\gamma^\mu t_L Z_\mu + \frac{g}{2c_W}\left(-\frac{4s_W^2}{3} + \kappa_R^{NC}\right)\overline{t_R}\gamma^\mu t_R Z_\mu$$

$$+ \frac{g}{\sqrt{2}}\left(1 + \kappa_L^{CC}\right)\overline{t_L}\gamma^\mu b_L W_\mu^+ + \frac{g}{\sqrt{2}}\left(1 + \kappa_L^{CC*}\right)\overline{b_L}\gamma^\mu t_L W_\mu^-$$

Note the appearance of the cutoff parameter Λ already mentioned in the introduction. It is not surprising that T (or equivalently $\epsilon_1$) and $\epsilon_b$ receive such contribution since we know that they are dominated by the top loop contributions and therefore not only depend on the top mass but also on the couplings of the gauge bosons to t/b.

One can easily interpret the two bands shown in figure 3. Indeed neglecting the quadratic terms and assuming that the charge current term does not deviate from the SM, one sees that $\epsilon_1$ essentially constrains $\kappa_R^{NC} - \kappa_L^{NC}$ that is to say the axial coupling of Ztt which corresponds to the horizontal band. $\epsilon_b$ is mainly sensitive to $\kappa_L^{NC}$ the left handed component which goes like a V-A component and corresponds to the diagonal band.

In the following I will arbitrarily (but with some quantitative arguments given below for the LH case) assume that the compensating contributions are such that $|\delta\epsilon_1/\epsilon_1|<1.5$ and $|\delta\epsilon_b/\epsilon_b|<1.5$ (with $\epsilon_1=4.8\cdot10^{-3}$ and $\epsilon_b=-6.5\cdot10^{-3}$ in the SM). Experimentally one can in principle constrain $\delta\epsilon_b$ from Zbb LEP1 measurements but, as already noted, these measurements indicate significant deviations which limits this type of constraint. I also will assume that Λ=1 TeV which gives the loosest constraint on couplings.

### 3.2.2 Gauge invariance

From gauge invariance one can write [17]:
$$\kappa_{bL}^{NC} + \kappa_{tL}^{NC} \sim \kappa_{tL}^{NC} = 2\kappa_{tLbL}^{CC}$$
In terms of our conventions:
$$\frac{\delta WtLbL}{WtLbL} = 0.72 \frac{\delta ZtLtL}{ZtLtL}$$

→ ZtLtL is directly related to the charged current CC but CC is so far poorly constrained by single top measurement (see section 5.1)

One can therefore consider that this relation gives for the various models a **prediction** for $\frac{\delta WtLbL}{WtLbL}$ which could be tested when the single top measurement will become more precise.

### 3.2.3 B physics constraints

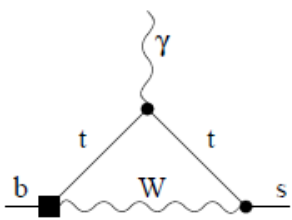

This process receives a Wtb contribution which allows deriving some constraints on this coupling. It turns out however that this constraint overwhelmingly concerns WtRbR which therefore screens eventual a BSM WtLbL. Adding to the previous Lagrangian the term:

$$\frac{g}{\sqrt{2}}\kappa_R^{CC}\overline{t_R}\gamma^\mu b_R W_\mu^+ + \frac{g}{\sqrt{2}}\kappa_R^{CC*}\overline{b_R}\gamma^\mu t_R W_\mu^-$$

As defined in [18], one has:
$$\mathcal{B} = (3.15 \pm 0.23) - 8.2\kappa_L + 427\kappa_R$$
and, given that the experimental value, is compatible with the SM prediction, this means that the right-handed coupling WtRbR is constrained, within 2 sd, at 2*0.23/427~1% assuming that there is no significant compensation from $\kappa_L$.

Similarly the process Bs->µµ, which goes through loops involving ZtLtL (penguins) and WtLbL (box) couplings, could be affected by some of the predictions of table I. This branching ratio goes like V²tb, a factor in common to the two diagrams. At present CMS+LHCb give BR=(2.9±0.7)10$^{-9}$ with a prediction (3.35±0.28)10$^{-9}$ which gives Vtb=0.93±0.12. This result is however too naive since the penguin diagram, which is dominant, also depends on the ZtLtL coupling which is related to Vtb through gauge invariance.

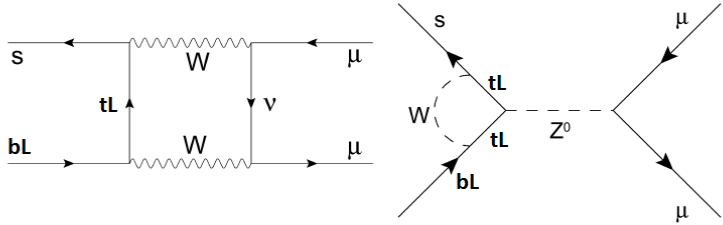

Mass mixing for the Bs mesons depends on a box diagram with top exchange and goes like Vtb².
The experimental accuracy is excellent but the theory uncertainty is at the level of 15%.
From the most recent update from LHCb, one can deduce that Vtb=1.01±0.0013±0.08(th) which gives a comparable accuracy to the single top measurement.

These examples show how B physics and linear collider physics are complementary for the tests of top EW couplings.

## IV Consequences for the proposed models

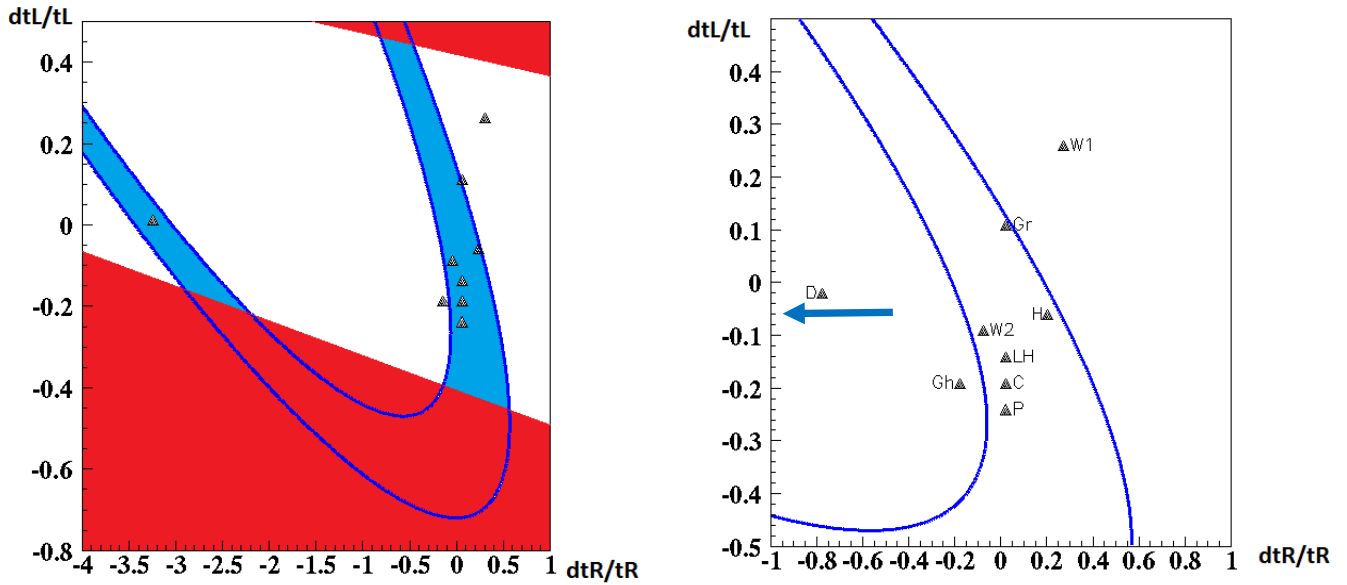

Figure 4: The allowed regions for $\epsilon_1$ are in between the two blue curves while the red regions are excluded by $\epsilon_b$. The black triangles correspond to the various models which are defined according to initials. Note that solution D on the right-hand side has to be displaced to dtL/tL= -3.3

Most models are consistent with the loop constraints. The general trend is to expect a moderate relative variation on tL Z couplings. Due to non linear terms present in the expression of $\epsilon_1$, large relative variations are allowed for tR Z couplings but, with the exception of the Djouadi et al case commented below, are not predicted.

**Table I: Expected variations for tL and tR couplings to Z and tLbL to W for various models and the corresponding relative variations for $\epsilon_1$ and $\epsilon_b$. The effect for the cross section ttZ is also shown. In red are the numbers which could give a significant signal at LHC.**

| Model | dtR/tR % | dtL/tL % | dtLbL/tLbL % | d$\epsilon_b$/$\epsilon_b$ | d$\epsilon_1$/$\epsilon_1$ | d$\sigma$Ztt/$\sigma$Ztt % |
|---|---|---|---|---|---|---|
| Carena | 0 | -20 | **-14** | 0.8 | 1.1 | **-30** |
| Djouadi | -330 | 0 | 0 | -1.4 | 1.1 | **70** |
| Gherghetta | -20 | -20 | **-14** | 0.7 | 2.1 | **-36** |
| Grojean | 0 | 10 | 7 | -0.4 | -1.0 | 17 |
| Hosotani | 18 | -7 | -5 | -0.4 | -0.8 | -5 |
| Little Higgs | 0 | -15 | -10 | 0.6 | 0.9 | -23 |
| Pomarol | 0 | -25 | **-17** | 1.0 | 1.2 | **-37** |
| Wulzer 1 | 25 | 25 | **17** | -1.1 | 5.8 | **56** |
| Wulzer 2 | -10 | -10 | -7 | 0.4 | 1.3 | -20 |

### 4.1 Comments on the Djouadi et al model

The Djouadi et al model does not really predict ZtRtR but allows a large interval which can be restricted by the oblique contributions. In reference [7] are given the various definitions needed for the following discussion. The value dbR/bR=0.2 deduced from LEP1 can be related to dtR/tR by [19]:

$\frac{dt_R}{t_R} / \frac{db_R}{b_R} \sim \frac{Q_b I_{3R}^{tR}}{Q_t I_{3R}^{bR}} \frac{F(c_{tR})}{F(c_{bR})}$ where Qb=-1/3 and Qt=2/3 while the ratio of isospins is model dependent and will be taken as 1 to satisfy the constraints. Indeed, as can be seen on figure 4, if this ratio was -1, dtR/tR would be large and positive which, even for non vanishing values of dtL/tL, cannot satisfy loop constraints.

Reference [7] gives F(cbR)~-0.25. To avoid having light KK quarks which are already excluded by LHC, one should restrict to ctR>-0.5 meaning that dtR/tR varies typically between -0.8 and -3.5 (hence the arrow indicated in figure 4). For dtR/tR=-0.8 one can see from figure 4 that dtL/tL needs to be very negative to belong to the allowed region which does not fall naturally within the model under consideration. On the contrary if one chooses dtR/tR=-3.2, this solution falls in the middle of the allowed $\epsilon_1$ band and this will therefore be retained as the Djouadi et al. solution (see appendix for more details).

For this solution ZtLtL, ZbLbL and (through gauge conservation) WtLbL do not receive appreciable BSM contributions which seems to imply that tL and bL are not 'composite-like'. This argument is however a bit short-sighted since, as already mentioned, the absence of ZbLbL could be due to a symmetry cancellation which does not occur for ZtLtL and WtLbL. Recalling that this model only invokes Z-Z' mixing, it is perfectly plausible to assume that there are other sources, like tL mixing with heavy quarks, which could affect ZtLtL and WtLbL. This possibility is consistent with the RS model which predicts the existence of heavy quarks which could mix with tL.

In the appendix, I show that a Djouadi solution can be fully solved using ILC potential to separate $\gamma$ from Z contributions. Using energy dependence one can even determine the Z' mass with a 4% accuracy.

## 4.2 Comments on the LH model

The Little Higgs model (LH) has a wide range of predictions for dtL/tL with dtR/tR=0 and I have retained a kind of average value which agrees with constraints. In this model [20], the compensating terms can be computed.

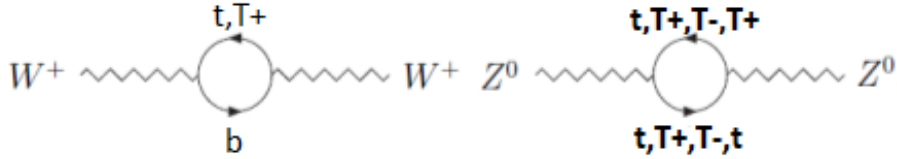

For what concerns $\epsilon_1$, one needs negative compensating contribution which can be provided by this model (see diagrams above and reference [21]).

For what concerns $\epsilon_b$, there is a loop contribution from the heavy quarks T which gives a compensation provided that T has a mass below 1 TeV.
From these 2 examples, one retains that constraints are very useful to guide an educated guess and considerably reduce the domain of parameters allowed by the proposed models. The presence of compensating terms of an adequate size seems to occur naturally either from the heavy quark sector or from other sources like, for instance, from the Higgs couplings effects discussed in chapter II.

# V Prospects for colliders

## 5.1 LHC

Most of the models corresponding to figure 2 predict heavy quarks which could be discovered at LHC14. This is also true for heavy vector bosons but, as already mentioned, these discoveries rely on colored vector bosons which could be difficult to observe since they primarily decay into energetic top jets which are difficult to reconstruct at LHC (the 'boosted top' problem). In case they are heavy enough to decay into the predicted heavy quarks their observation in top pairs is even more difficult since one predicts a very large total width with $\Gamma/M\sim1$.

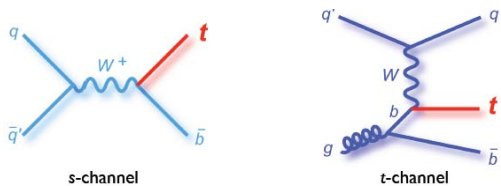

| experiment | Vtb |
|---|---|
| CDF | 0.92+0.10-0.08 |
| D0 | 1.12+0.09-0.08 |
| CMS | 1.03+-0.12±0.04(th) |
| ATLAS | 1.04+.10-0.11 |

Single top production measurements allow to access to the Wtb coupling which is usually expressed in terms of Vtb (see for instance [22]). Tevatron will soon publish a measurement combining D0 and CDF data with an error below 8%. The table above summarizes the present situation. By naively averaging, one gets Vtb~1.03±0.05±0.04(th) which already allows to make a first selection among the predictions of Table I.

For what concerns the ttZ cross section, recall the present CMS result on the basis of 5 fb-1:
$\sigma_{t\bar{t}Z} = 0.28^{+0.14}_{-0.11}(stat.)^{+0.06}_{-0.03}(syst.) pb$ while the NLO prediction gives $\sigma_{t\bar{t}Z} = 0.137^{+0.012}_{-0.016} pb$

Assuming, crudely speaking, that the statistical error can be divided by 5 and the systematics by 2 one can reasonably expect an error of 13% on this quantity. This hypothesis has been used for Table II. Note the big asymmetry for what concerns the error on tR which (as for bR at LEP1) takes into account the sign ambiguity at LHC which does not occur at ILC given the γ−Z interference.

ATLAS and CMS will be able to test several predictions given in Table I. In the last column of Table I, I have indicated the expected variation on the ttZ cross section. Four models predict an observable deviation in ttZ at LHC. Djouadi et al. predicts the largest deviation already measurable with present data. While this measurement provides no insight on the origin of a deviation, as due to tL or tR couplings, it is not excluded that the origin could be inferred by measuring the decay distributions of the top quarks accompanying the Z boson[19]. One could also combine this measurement with the single top result which allows, through gauge invariance, to infer the tL Z coupling and, accordingly, deduce the tR contribution to the ttZ cross section.

## 5.2 ILC

At ILC, all top EW couplings, including Wtb, will be measured at a % allowing full separation between axial and and vectorial couplings and between Ztt and γtt couplings. This analysis is described in [23] and figure 5 recalls these performances. One can (see Table II), alternatively, cast these results into tL and tR couplings recalling that ILC allows a perfect separation of these two contributions. Note that for this table there is no assumption about the photon couplings, meaning that the right and left handed couplings are determined independently. If one assumes that the photon cannot acquire an axial component then this constraint reduces the error on these couplings to 0.25% (instead of 0.6%). In the RS example given in the appendix these two couplings vary independently.

For the Wtb coupling one can use the top total width directly accessible from the threshold scan [24]. Given that this width measurement can be done at the 1.5% level one can deduce the WbLtL coupling at the 0.75% level. Deviations could also come from the BSM WtRbR couplings but recall that the b->sγ measurement sets a strict limit on WbRtR at ~0.1% therefore restricting this interpretation.

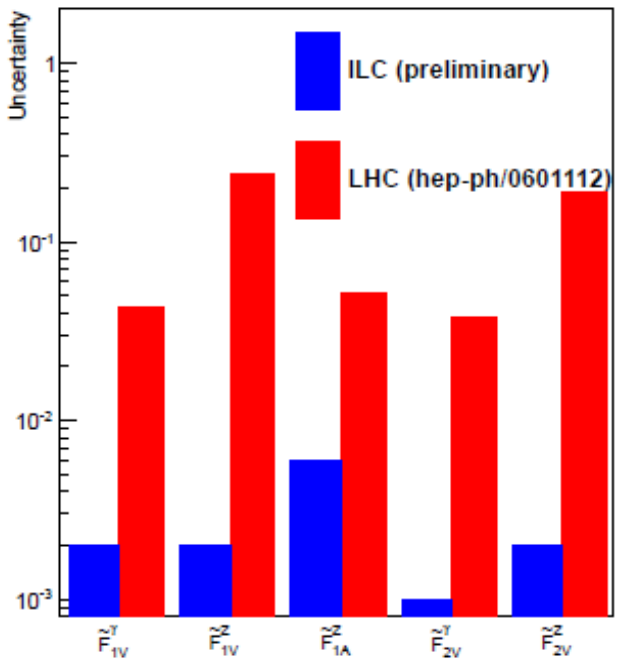

|  | ILC | LHC σttZ | |
|---|---|---|---|
| δZtLtL/ZtLtL % | 0.6 | ±8 | |
| δZtRtR/ZtRtR % | 1.4 | -240 | 40 |
| δγtLtL/γtLtL % | 0.6 | -7 | 12 |
| δγtRtR/γtRtR % | 0.6 | -7 | 12 |

Table II: ILC and LHC expected accuracies in terms of the tL and tR couplings

**Figure 5:** Comparison of statistical precisions on CP conserving axial and vector form factors expected at LHC[15] and ILC[23]. The LHC results assume an integrated luminosity of L = 300 fb-1. The results for ILC assume an integrated luminosity of L = 500 fb-1 at 500 GeV and polarized beams.

### 5.2.1 Higgs and top couplings measurements

The top EW measurements are complementary to the Higgs coupling measurements described in section II. In composite models it is necessary to combine eventual deviations on Higgs and top EW couplings to fully interpret oblique contributions. In the example on Higgs coupling deviations given in section II, I was assuming that a compensating positive contribution of 0.2 on the T variable would be needed. This is equivalent to have $\delta\epsilon_1/\epsilon_1$=0.35% which, see Table I, can be provided in many of the composite-like models.

Finally one may ask the following question: given the potentially large contributions to $\epsilon_1$ predicted in Table I, how is it that the oblique contributions measured at LEP1/SLC come so close to the SM values? This feature seems to imply that either there are strong cancellations between contributions of various origin, including from the Higgs sector, or that one should expect minute effects from BSM physics which therefore will require the highest accuracies to become significant.

## VI Conclusion

The lesson from this note is that, given a BSM model which predicts deviations on top couplings to Z, loop+gauge constraints allow to test the consistency of this model at the quantum level and, in some cases, restrict or even exclude these predictions. Some of them require compensating loop contributions and I could show, for the Little Higgs model, that they could be present if the new heavy quarks have masses ~1 TeV which could be discovered at LHC.

At LHC, one can measure indirect signals through single top production or ttZ cross section measurements. B physics can also provide important inputs about ZtLtL couplings.

This type of arguments can also be used for Higgs couplings and, in case there are no direct signals observed at LHC, one can still hope to learn a lot about BSM physics from the deviations observed on Higgs and top EW couplings. Both sectors are therefore complementary and obey to similar constraints derived from LEP/SLC measurements. While one cannot exclude the presence of large coupling deviations observable at LHC, it might require ILC accuracies to access to fully significant deviations.

### Acknowledgements

 I would like to thank Gino Isidori and Emi Kou for providing their expertise on B physics. Thanks to Roman Poeschl for a critical reading of this note.

### References

**[1]A Large mass hierarchy from a small extra dimension**
Lisa Randall (Princeton U. & MIT, LNS), Raman Sundrum (Boston U.). May 1999.
Published in **Phys.Rev.Lett. 83 (1999) 3370-3373**
e-Print: **hep-ph/9905221**

**[2]Little Higgs review**
Martin Schmaltz (Boston U.), David Tucker-Smith (Williams Coll.). Feb 2005.
Published in **Ann.Rev.Nucl.Part.Sci. 55 (2005) 229-270**
e-Print: **hep-ph/0502182**

**[3]Conference**: LHC Run1 Aftermath 2013, Where Theory Meets Experiment
**Title**: Status and prospects of the electroweak fit of the SM with Gfitter after the Higgs discovery
**Speaker**: R. Kogler (University of Hamburg)
http://project-gfitter.web.cern.ch/project-gfitter/

**[4]Strong Higgs Interactions at a Linear Collider.**
Roberto Contino, Christophe Grojean, Duccio Pappadopulo, Riccardo Rattazzi, Andrea Thamm, . Sep 2013.
e-Print: **arXiv:1309.7038**

**[5]Higgs Couplings and Electroweak Phase Transition**
Andrey Katz, Maxim Perelstein. Jan 8, 2014. 27 pp.
e-Print: **arXiv:1401.1827**

**[6]Light Kaluza Klein States in Randall-Sundrum Models with Custodial SU(2)**
Marcela S. Carena (Fermilab), Eduardo Ponton (Columbia U.), Jose Santiago (Fermilab), Carlos E.M. Wagner (Argonne, HEP & Chicago U., EFI & Chicago U., KICP). Jul 2006.
e-Print: **hep-ph/0607106**

**[7]** (Djouadi) **Resolving the A(FB)**b puzzle in an extra dimensional model with an extended gauge structure**
Abdelhak Djouadi, Gregory Moreau (Orsay, LPT), Francois Richard (Orsay, LAL). Oct 2006.
Published in **Nucl.Phys. B773 (2007) 43-64**
e-Print: **hep-ph/0610173**

**[8]Fermion Masses in Emergent Electroweak Symmetry Breaking**
Yanou Cui (Harvard U., Phys. Dept.), Tony Gherghetta, James Stokes (Melbourne U.). Jun 2010.
Published in **JHEP 1012 (2010) 075**
e-Print: **arXiv:1006.3322**

**[9]Light top partners and precision physics**
Christophe Grojean (Barcelona, IFAE & CERN), Oleksii Matsedonskyi (INFN, Padua), Giuliano Panico (CERN). Jun 19, 2013.
e-Print: arXiv:1306.4655

**[10]Higgs boson mass and electroweak-gravity hierarchy from dynamical gauge-Higgs unification in the warped spacetime**
Yutaka Hosotani, Mitsuru Mabe (Osaka U.). Mar 2005. 1
Published in **Phys.Lett. B615 (2005) 257-265**
e-Print: **hep-ph/0503020**

**[11]Top quark properties in little Higgs models**
C.F. Berger (SLAC), M. Perelstein (Cornell U., CIHEP), F. Petriello (Wisconsin U., Madison). Dec 2005.
e-Print: **hep-ph/0512053**

**[12]Top Quark Compositeness: Feasibility and Implications**
Alex Pomarol, Javi Serra (Barcelona, IFAE). Jun 2008.
Published in **Phys.Rev. D78 (2008) 074026**
e-Print: **arXiv:0806.3247**

**[13]**A. Wulzer, O.Matsedonskyi and G.Panico, **private communication**

**[14]A Custodial symmetry for Zb anti-b.**
Kaustubh Agashe, (Syracuse U.) , Roberto Contino, (Rome U. & INFN, Rome & Johns Hopkins U.) , Leandro Da Rold, (Barcelona, IFAE) , Alex Pomarol, (Barcelona, IFAE & CERN) Published in **Phys.Lett.B641:62-66,2006**.
e-Print: **hep-ph/0605341**

**[15] Report of the 2005 Snowmass top/QCD working group.**
A. Juste *et al.*
e-Print: **hep-ph/0601112**

**Probing electroweak top quark couplings at hadron colliders**
U. Baur (SUNY, Buffalo), A. Juste (Fermilab), L.H. Orr, D. Rainwater (Rochester U.). Nov 2004.
Published in **Phys.Rev. D71 (2005) 054013**
e-Print: **hep-ph/0412021**

**[16] Analysis of $tbW$ and $ttZ$ couplings from CLEO and LEP / SLC data**
F. Larios (Merida, IPN), M.A. Perez (CINVESTAV, IPN), C.P. Yuan (Michigan State U.). Mar 1999. 11 pp.
Published in **Phys.Lett. B457 (1999) 334-340**
e-Print: **hep-ph/9903394**

**[17] Top couplings and top partners**
J.A. Aguilar-Saavedra, M. Pérez-Victoria (CAFPE, Granada & Granada U., Theor. Phys. Astrophys.). Feb 22, 2013.
Published in **J.Phys.Conf.Ser. 452 (2013) 012037**
e-Print: **arXiv:1302.5634**

**[18] Anomalous Wtb coupling effects in the weak radiative B-meson decay**
Bohdan Grzadkowski, Mikolaj Misiak (Warsaw U. & CERN). Feb 2008.
Published in **Phys.Rev. D78 (2008) 077501, Erratum-ibid. D84 (2011) 059903**
e-Print: **arXiv:0802.1413**

**[19] Can LHC observe an anomaly in tt̄Z production?**
François Richard (Orsay, LAL). Apr 12, 2013.
e-Print: **arXiv:1304.3594**

**[20] Electroweak precision constraints on the littlest Higgs model with T parity.**
Jay Hubisz, Patrick Meade, Andrew Noble, Maxim Perelstein, (Cornell U., LEPP) . Jun 2005. 22pp.
Published in **JHEP 0601:135,2006**.
e-Print: **hep-ph/0506042**

**[21] Updated Status of the Global Electroweak Fit and Constraints on New Physics**
M. Baak (CERN), M. Goebel (DESY & Hamburg U.), J. Haller (Hamburg U. & Gottingen U., Zweites Phys. Inst.), A. Hoecker (CERN), D. Ludwig (DESY & Hamburg U.), K. Moenig (DESY), M. Schott (CERN), J. Stelzer (Michigan State U.). Jul 2011.
Published in **Eur.Phys.J. C72 (2012) 2003**
e-Print: **arXiv:1107.0975**

**[22] Observation of the associated production of a single top quark and a W boson in pp collisions at √S = 8 TeV**
CMS Collaboration (Serguei Chatrchyan (Yerevan Phys. Inst.) *et al.*). Jan 13, 2014.
CMS-TOP-12-040, CERN-PH-EP-2013-237
e-Print: **arXiv:1401.2942**

**[23] A precise determination of top quark electro-weak couplings at the ILC operating at 500 GeV**
M.S. Amjad, M. Boronat, T. Frisson, I.Garcia Garcia, R. Poschl, E. Ros, F. Richard, J. Rouene, P.Ruiz Femenia, M. Vos, . IFIC-13-06, LAL-13-265, Jul 2013. Temporary entry
e-Print: arXiv:1307.8102

**[24] Study of top quark pair production near threshold at the ILC**
Tomohiro Horiguchi, Akimasa Ishikawa, Taikan Suehara, Keisuke Fujii, Yukinari Sumino, Yuichiro Kiyo, Hitoshi Yamamoto. Oct 2, 2013.
e-Print: **arXiv:1310.0563**

# APPENDIX: an example with RS

In this model [7] one assumes that there is Z-Z' mixing, Z' being an extra Z boson with no KK zero mode and no direct coupling to light fermions. Z' is dominantly coupled to bR and tR. Its coupling to bR explains the deviation observed at LEP1 on AFBb. As explained in section 4.1, one expects a much stronger effect on tR while one can almost neglect the effect on tL. For completeness the formulae below show not only the mixing terms but also the propagator terms due to $\gamma$ and Z KK excitations. This fairly complex model, allowed within RS, is taken as an example of what ILC can do to disentangle a complex mechanism. The defrinitions of the terms used below can be found in ref [7].

$$\frac{dR_Z}{R_Z} = \left(\frac{M_Z}{0.4 M_{KK}}\right)^2 \left[1 + \frac{\frac{3}{4}\left(1 - \frac{4}{3}\sin^2\theta'\right)}{\sin^2\theta' \cos^2\theta'}\right] F(c_{tR}) + \frac{s}{s - M_{KK}^2} Q(e) Q(c_{tR})$$

$$\frac{dL_Z}{L_Z} = \left(\frac{M_Z}{0.4 M_{KK}}\right)^2 \left[1 - \frac{1}{4\cos^2\theta'}\right] F(c_{tL}) + \frac{s}{s - M_{KK}^2} Q(e) Q(c_{tL})$$

$$\frac{dR_\gamma}{R_\gamma} = \frac{s}{s - M_{\gamma KK}^2} Q(e) Q(c_{tR})$$

$$\frac{dL_\gamma}{L_\gamma} = \frac{s}{s - M_{\gamma KK}^2} Q(e) Q(c_{tL})$$

F(c) and Q(c) are simply related by Q(c)=-0.8F(c)+0.2. Given the low electron mass Q(e) should be at its minimum value 0.2 and therefore one has 6 unknowns: Mkk (assuming the same mass for KK excitations for photons Z and Z'), sinθ', Q(ctR) Q(ctL) and sin²θL/R and 4 observables which can be measured at two energies (0.5 TeV and 1 TeV for ILC) allowing full disentangling with redundancy. From reference [7] one takes Mkk=3 TeV sinθ'=0.1 and , as explained in the text, one predicts F(tR)=-6.9. Note in passing that in ref. [7] one has F(bR)=-0.25 meaning that the ratio F(tR)/F(bR)~30 pretty close to mt/mb as one would expect in the RS scheme.

**ZtLtL**
From [7] and assuming weak isospin symmetry, F(bL)=F(tL)=-2.1 Q(tL)=1.85. From the later one deduces that dtL/tL=-2% which is very small, barely measurable at ILC. In addition one may have contributions due to tL mixing with heavy quarks. Figure 4 tells us that for the D solution this contribution cannot exceed ±10% as is the case for most of the models and also for reference [17] which takes into account existing constraints. A mixing term between tL and a vector-like quark T which belongs to a isospin singlet gives a contribution $-\sin^2\theta_L$ to the left-handed terms. This quark mixing term, as explained in [17], is tighly constrained by LEP/SLC measurements and mass limits on heavy quarks from LHC, meaning that $\sin^2\theta_L$ <0.03.

**Mkk**
To extract Mkk one can look at the energy dependence of dRγ/Rγ which has no bosonic mixing. Taking the ratio of dRγ/Rγ at 0. 5 and 1 TeV and assuming the same statistical accuracy one has:

$$\frac{dM_{KK}}{M_{KK}} = \pm 4\%$$

This method could be affected by quark mixing for tR. As suggested by the figure below, one can therefore extrapolate the energy dependence of dRγ/Rγ to s=0. Indeed one expects an approximately straight line for s<<M²kk passing through the origin if there is no quark mixing.

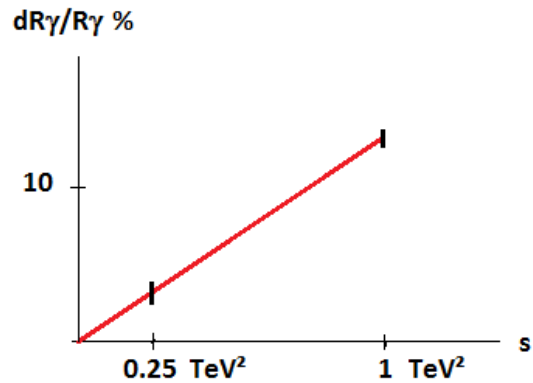

**RS at LHC**

As explained in the main text, the large deviation expected on tR couplings can be deduced from the ttZ cross section measurement at LHC. For what concerns the photon couplings, however, this model predicts zero effect given that at LHC the photon is on mass shell and above formulas show no deviation on photon couplings for s=0. When the top quark mixes with a heavy quark this could affect the photon couplings but note that the effect is expected to be small and presumably outside the reach of LHC.